# Fractional order [PI] Controller and Smith-like Predictor Design for A Class of High Order Systems

Zhenlong Wu[1], Jie Yuan[2], Yuquan Chen[3], Donghai Li[1], YangQuan Chen[4*]

1. State Key Lab of Power Systems, Department of Energy and Power Engineering, Tsinghua University, Beijing 100084, PR China

2. School of Automation, Southeast University, Nanjing 210096, China.

3. Department of Automation, University of Science and Technology of China, Hefei, 230026, China

4. Systems and Automation (MESA) Lab, School of Engineering, University of California, Merced, CA 95343, USA

(Corresponding author e-mail: ychen53@ucmerced.edu)

**Abstract:** To handle the control difficulties caused by high-order dynamics, a control structure based on fractional order [proportional integral] (PI) controller and fractional order Smith-like predictor for a class of high order systems in the type of $K/(Ts+1)^n$ is proposed in this paper. The analysis of the tracking and disturbance rejection is illustrated based on the terminal value theorem and shows that the proposed control structure can ensure that the closed-loop system converges to the set point without static error and the closed-loop system recovers to its original state when the input disturbance occurs. Then, simulations about the influence on the control performance and control signal with different $\chi$ are carried out based on multi-objective genetic algorithm (MO-GA). The results show that the control performance can be improved and the energy of the control signal can be reduced simultaneously when the order $\chi$ is chosen no more than one. This can verify that the fractional order Smith-like predictor with $\chi < 1$ has an advantage over that of the integral order Smith-like predictor with $\chi = 1$.

**Key Words:** Fractional order [PI] Controller, fractional order Smith-like predictor, high order systems, multi-objective genetic algorithm

## 1  INTRODUCTION

Heat transfer and fluid flow are common phenomena in chemical and energy processes [1]. These processes contain lots of energy transfer and material conversion. Their dynamics are typical distributed parameter systems and often have slow responses to disturbances and set points because of the large scale and complexity in actual industrial system [2]. To better characterize the dynamics of those systems and more easy design the controller, a class high order systems with the transfer function in the form of $K/(Ts+1)^n$ is often identified [3 - 4]. Note that $K$, $T$ and $n$ are the gain, the time constant and the order of the high order system, respectively, and we have $n \geq 3$. Superheated steam temperature system and main steam pressure system are these typical high order systems. Note that the order of the actual system is exactly unknown and may be non-integer [2]. There are some control difficulties caused by high-order dynamics, such as slow response speed and unknown accurate mathematical models based on the mechanism.

To improve the control performance of the high order systems, special tuning rules of proportional integral (PI) controller are developed for these high order systems [5]. Besides, some control structures are also proposed to enhance the tracking performance and disturbance rejection such as Smith predictor (SP) and Internal model control (IMC) by predicting and compensating dead time for first order plus dead time (FOPDT) systems or second order plus dead-time (SOPDT) systems with the help of model reduction methods. Recently, a modified active disturbance rejection control (ADRC) is proposed to handle the slow response caused by high-order dynamics [4]. However, the modified control structure does not consider the influence of the actual systems' order on the control performance which may impair the performance of the closed-loop system.

In the past decades, the fractional calculus has experienced an explosive development [6 - 7]. The fractional order systems such as the gas turbine system [8] and perturbed pressurized heavy water reactor system [9] show more dynamic information of the complicated systems than that of integer order systems. Fractional order PI controller as a generalization of integer order PI controller has greater flexibility and shows better control performance than integer order PI controller [10]. With the development of the fractional order control theory, fractional order PI controller with the type of $PI^\lambda$ and $[PI]/(PI)^\lambda$ have attracted many attentions [11 - 13]. The latter can outperform $PI^\lambda$ controller with the control specifications which is the interesting point of research in this paper [13]. A control structure with $(PI)^\lambda$ controller as the feedback controller and fractional order Smith-like predictor is proposed for the higher order system mentioned above in

This work is supported by China Scholarship Council (CSC) under Grant 201806210219.

this paper. How to design the fractional order Smith-like predictor is the key research and the influence of the fractional order choice on control performance is studied carefully. The main contributions of this pare are summarized as:

1) A control structure based on $(PI)^\lambda$ controller and a fractional order Smith-like predictor are proposed to compensate the high-order dynamics for a class of high order systems.

2) The tracking and disturbance rejection performance of the proposed control structure are analyzed.

3) The influence analysis of the fractional order of Smith-like predictor on the control performance with the help of multi-objective genetic algorithm (MO-GA) is carried out. The results show that the control performance can be improved and the energy of the control signal can be reduced simultaneously when the order is chosen no more than one.

This paper comprises of six sections and the rest of the paper is organized as follows. Section 2 formulates the proposed control structure. The analysis about the tracking and disturbance rejection and the research objectives are presented in Section 3. The analysis about the influence on the control performance and control signal with different orders of Smith-like predictor are discussed in Section 4. The necessary discussions and concluding remarks are shown in Section 5 and Section 6, respectively.

## 2 THE PROPOSED CONTROL STRUCTURE

Consider a high order system depicted by

$$G = \frac{K}{(Ts+1)^n}, \quad (1)$$

where $K$, $T$ and $n$ are the gain, the time constant and the order of a high order system, respectively, and $n \geq 3$. Note that the order of the actual system may be non-integer. The proposed control structure can be shown in Fig. 1. $(PI)^\lambda$ controller roles as the feedback controller which is depicted by

$$G_c = \left(k_p + \frac{k_i}{s}\right)^\lambda, \quad (2)$$

where $k_p$, $k_i$ and $\lambda$ are the proportional gain, integral gain and the order of FO [PI] controller, $\lambda \in (0, 2)$. The high order in Equation (1) is divided into two parts depicted by

$$G_{p1} = \frac{K}{(Ts+1)^\chi} \quad (3)$$

and

$$G_{p2} = \frac{1}{(Ts+1)^{(n-\chi)}}. \quad (4)$$

Compared with the regular control structure of SP as shown in Fig. 2, $G_{p1}$ and $G_{p2}$ role as the functions of the delay-free part of the system ($G_{mo}$) and the delay time part ($e^{-\tau s}$) in SP, respectively.

The objective of SP is to eliminate adverse effects of delay time by the predicted control structure in Fig. 2 when the delay time is known exactly. The similar objective is to eliminate adverse effects caused by the high-order dynamics for the proposed control structure no matter the order of the actual system is known exactly. Therefore, we choose $\chi \in (0, 2)$ considering that there will has more high-order dynamics which cannot be predicted and eliminated if the order $\chi$ is larger than two.

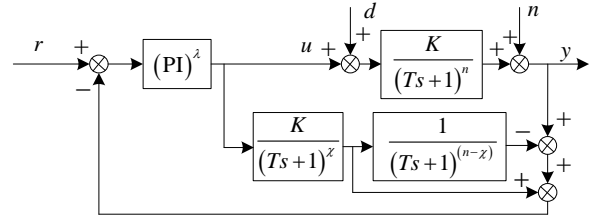

Fig. 1. The proposed fractional order smith-like control structure.

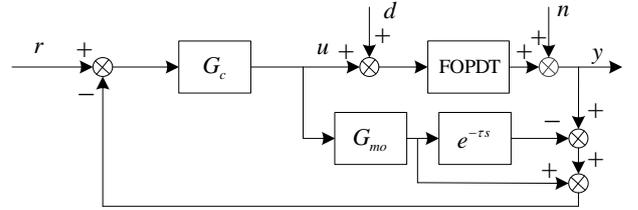

Fig. 2. The regular control structure of Smith predictor.

*Remark*: Note that $(PI)^\lambda$ controller and $G_{p1}$ can be implemented by the approximate method of frequency domain response in Ref [14]. The code can be downloaded from MATALB central or obtained by emailing to the corresponding author.

## 3 The NECESSARY ANALYSIS AND RESEARCH OBJECTIVE

In fact, the proposed control structure can be reorganized in the framework of a feedback control structure as shown Fig. 3 and $G_{ec}$ is the equivalent structure depicted by

$$G_{ec}(s) = \frac{G_c(s)}{1+G_c(s)G_3(s)} = \frac{G_c(s)}{1+G_c(s)G_{p1}(s)(1-G_{p2}(s))}, \quad (5)$$

where $G_c(s)$, $G_{p1}(s)$ and $G_{p2}(s)$ can be seen in Equations (2) - (4).

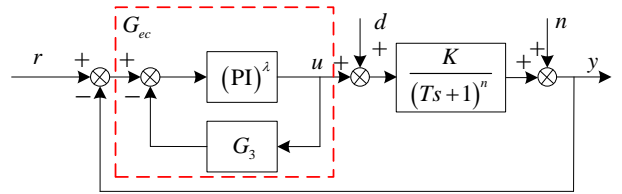

Fig. 3. The equivalent structure of the proposed control structure.

Based on the equivalent structure in Fig. 3, the transfer function from the set point $r$ to the output $y$ can be obtained as,

$$G_{yr}(s) = \frac{G_{ec}(s)G(s)}{1+G_{ec}(s)G(s)}. \quad (6)$$

Considering a step change in the set point ($k_1/s$) and combining with terminal value theorem, we can obtain

$$y(t)\big|_{t\to\infty} = \lim_{s\to 0} s \frac{G_{ec}(s)G(s)}{1+G_{ec}(s)G(s)} \frac{k_1}{s}$$

$$= \lim_{s\to 0} k_1 \frac{(k_p s + k_i)^\lambda}{s^\lambda (Ts+1)^n + (k_p s + k_i)^\lambda K(Ts+1)^{n-\chi}}$$

$$= k_1, \quad (7)$$

where $k_p$, $k_i$ and $k_1$ are the proportional gain, integral gain and the amplitude of the set point, respectively. Equation (7) verifies that the proposed control structure can ensure the closed-loop system converges to the set point without static error.

The transfer function from the input disturbance $d$ to the output $y$ is depicted by

$$G_{yd}(s) = \frac{G(s)}{1+G_{ec}(s)G(s)}. \quad (8)$$

Considering a step change in the input disturbance ($k_2/s$) and combining with terminal value theorem, we can obtain

$$y(t)\big|_{t\to\infty} = \lim_{s\to 0} s \frac{G(s)}{1+G_{ec}(s)G(s)} \frac{k_2}{s}$$

$$= \lim_{s\to 0} \frac{k_2 k}{(Ts+1)^n} \frac{(Ts+1)^n s^\lambda + (k_p s + k_i)^\lambda K\left((Ts+1)^{n-\chi}-1\right)}{(Ts+1)^n s^\lambda + (k_p s + k_i)^\lambda K(Ts+1)^{n-\chi}}$$

$$= 0, \quad (9)$$

where $k_p$, $k_i$ and $k_2$ are the proportional gain, integral gain and the amplitude of the input disturbance, respectively. Equation (9) demonstrates that the proposed control structure can ensure the closed-loop system recovers to its original state when the input disturbance occurs.

Based on the proposed control structure, the main research objective of this paper is about the influence analysis of the fractional order $\chi$ on the control performance and the energy of control signal with the help of MO-GA, and the range of $\chi$ for the controller synthesis is discussed.

MO-GA applied in this paper is a practical evolutionary algorithm which can be seen as a variant of NSGA-II [15]. MO-GA as an elitist GA always favors individuals with better fitness value (rank) and individuals that can help increase the diversity of the population even if they have a lower fitness value. By reducing the complexity of non-dominated sorting genetic algorithm, it can accelerate the speed of operation and convergence. MO-GA helps decision makers to select the appropriate Pareto solution by offering the Pareto optimal solution set of multiple objectives ($J_1$, $J_2$ …). As shown in Fig. 4, the solution set outside the Pareto front (the solution set shown in the blue circle in the figure) is not the Pareto optimal solution.

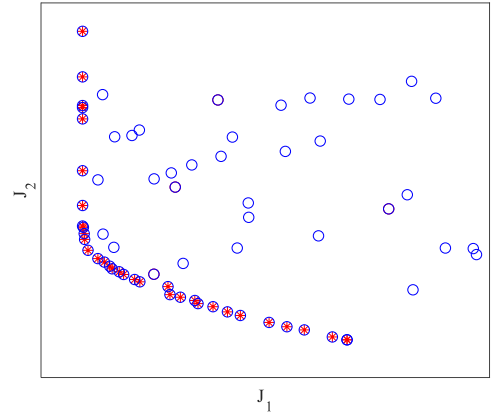

Fig. 4. The diagram of Pareto front.

The indices of the control performance and the energy of the control signal are chosen as multiple objectives ($J_1$, $J_2$) in this optimization. A commonly used metric as the control performance, the integrated time absolute error (ITAE) is given by

$$J_1 = \text{ITAE} = \int_0^\infty |r(t)-y(t)|t\,dt, \quad (10)$$

Note that the ITAE contains the indices of the tracking and disturbance rejection performance to better measure the control performance. Besides, the energy of the control signal is depicted by

$$J_2 = E_u = \int_0^\infty u^2(t)dt, \quad (11)$$

and the energy of the control signal is calculated during the whole simulation process. Note that $J_1$ and $J_2$ are conflicting objectives. A smaller $J_1$ means a strong control action that can result a larger $J_2$ and vice versa.

## 4 SIMULATIONS AND ANALYSIS

In this section, a typical high order system depicted by

$$G = \frac{1}{(20s+1)^4}, \quad (11)$$

is considered and the following simulations about the influence analysis of the fractional order $\chi$ on the control performance are carried out for the system in Equation (11). To better analyze the influence of the fractional order $\chi$ on the control performance, the different values of $\chi$ are set as 0.2, 0.4, 0.6, 0.8, 1.0, 1.2, 1.4, 1.6 and 1.8. Then the parameters of $(PI)^\lambda$ controller are optimized by MO-GA whose objectives are ITAE and $E_u$ as discussed in Section 3. The Pareto front with different $\chi$ are compared based on the optimized results.

Note that the following simulations are carried out based on *MATLAB* and *Simulink*. The solver type is "*fixed-step*" and the sample time is 0.0005s. For fair comparison, the same parameters of MO-GA are set for all different $\chi$: the population size is 30, the generations is 20, the crossover probability is 0.8, the recombination rate is 0.8, the crossover probability is 0.8, the Paretofraction is 0.35, and the mutation probability is 0.05.

The Pareto fronts with different $\chi$ are shown in Fig. 5. Note that the $J_1$ shown in the following figures is the actual value divided by one hundred. It can be seen that $J_1$ and $J_2$ locate in very large ranges. An extremely large $J_1$ or $J_2$ are both unreasonable for the engineering application, because that an extremely large $J_1$ means a bad control performance with slow tracking speed and weak disturbance rejection even though the energy of the control signal is small. Similarly, an extremely large $J_2$ means a large energy of the control signal which can result great challenges on actuators such as safety and wear, and the severely volatile control signal even though the total control performance is nice.

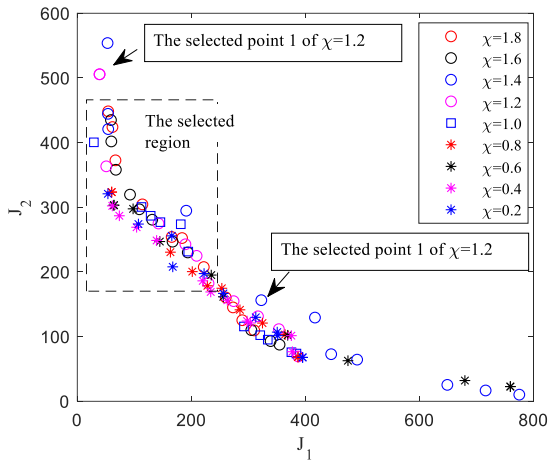

Fig. 5. The Pareto front with different $\chi$.

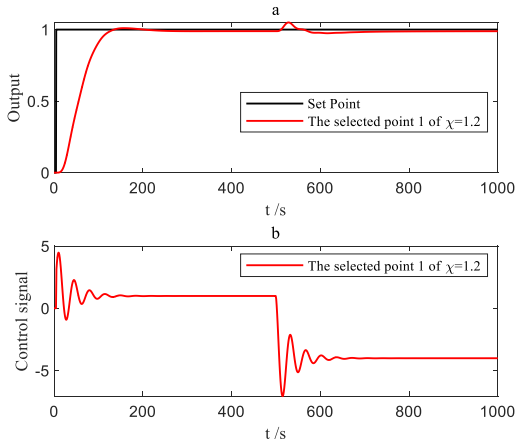

Fig. 6. The response of the selected point 1 of $\chi=1.2$. (a: the output response, b: the control signal)

To more intuitively explain the question discussed above, two pairs of {$J_1$, $J_2$} with $\chi=1.2$ are selected as shown in Fig. 5 which have an extremely large $J_1$ and $J_2$, respectively. The responses of this two pairs are shown in Fig. 6 and Fig. 7, respectively. The control signal with a large $J_2$ as shown in Fig. 6 is severely volatile even though the tracking and disturbance rejection performance are good. The response with a large $J_1$ as shown in Fig. 7 has a very slow tracking speed and needs long time to recover to the original state when the input disturbance occurs. Note that a unit set point changes at 5s and an input disturbance with a amplitude of 5 is added to the system at 500s in all simulations.

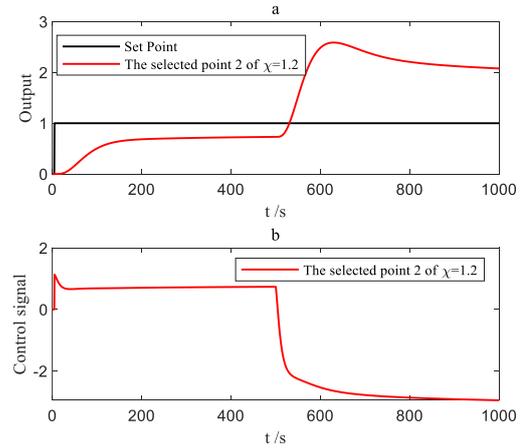

Fig. 7. The response of the selected point 2 of $\chi=1.2$. (a: the output response, b: the control signal)

Therefore, a reasonable region is selected as shown in the black dotted box of Fig. 5 which is a trade-off between $J_1$ and $J_2$. The local enlarged drawing of the selected region can be seen in Fig. 8. It can be seen that the situations of $\chi=0.2$, 0.4, 0.6, 0.8 ($\chi<1$) have better Pareto optimal solutions as shown in the cyan shadow of Fig. 8 which means that the control performance can be improved and the energy of the control signal is reduced simultaneously when the order $\chi$ is chosen no more than one. In a conclusion, the high order system with the proposed control structure can ensure the desired control performance by designing the proposed fractional order Smith-like predictor ($\chi<1$) and its superiority is verified.

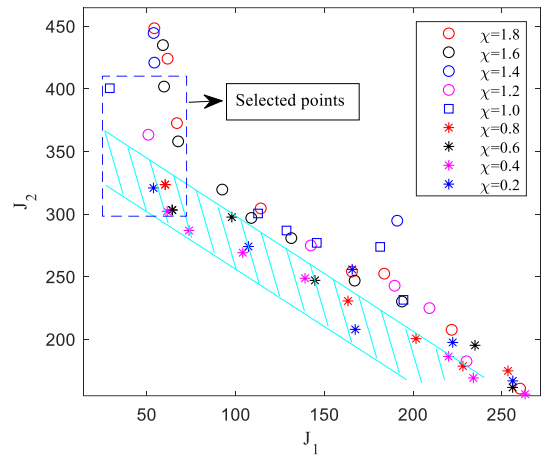

Fig. 8. The local enlarged drawing of the selected region.

We choose some typical pairs of {$J_1$, $J_2$} with different $\chi$ as the selected points in the black dotted box of Fig. 8. The responses of the closed loop system with these selected points are shown in Fig. 9 -Fig. 11 where Fig. 10 and Fig. 11 are the local enlarged drawings of the tracking performance and disturbance rejection, respectively. Note that there are two pairs of {$J_1$, $J_2$} with $\chi=1.6$ in the black dotted box of Fig. 8 named as $\chi=1.6(1)$ and $\chi=1.6(2)$ in Fig. 9 -Fig. 11.

When the order $\chi$ is larger than 1, the system has faster tracking speed than that of $\chi<1$ and the system also has worse disturbance rejection than that of $\chi<1$. What is more, the system with $\chi>1$ has a more serious fluctuation than that of $\chi=0.8$, 0.6, 0.4. The system with $\chi=0.2$ has the best disturbance rejection while its control signal from 500s to 1000s has a serious fluctuation. Besides, the system with $\chi=1.0$ has better disturbance rejection than that of other $\chi$ except $\chi=0.2$ while its control signal for the tracking and disturbance rejection both have a serious fluctuation.

## 5 DISCUSSIONS

The simulations in Section 4 show that the control performance can be improved and the energy of the control signal is reduced simultaneously when the order $\chi$ of the proposed structure is chosen no more than one. The main reason for the conclusion is that the proposed control structure with $\chi<1$ can predict and eliminate high-order dynamics as much as possible by the proposed fractional order Smith-like predictor. However, this does not mean a very small $\chi$ can result much better control performance. The excessive predictor and elimination cannot always obtain the improvement of the control performance. Note that the results in Fig. 9 cannot offer a good method to select an appropriate $\chi$ and how to select $\chi$ quantitatively will be the next work in future.

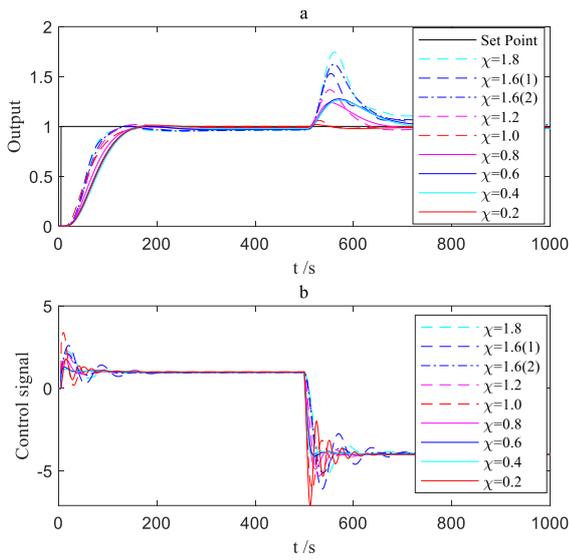

Fig. 9. The response of the select points in Fig. 8. (a: the output response, b: the control signal)

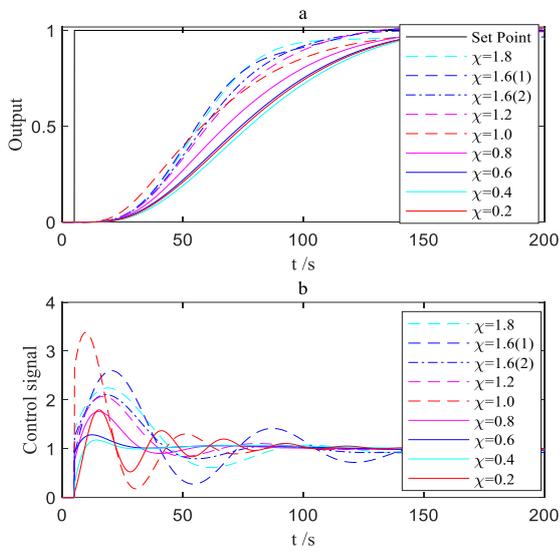

Fig. 10. The local enlarged drawing of the tracking performance in Fig. 9. (a: the output response, b: the control signal)

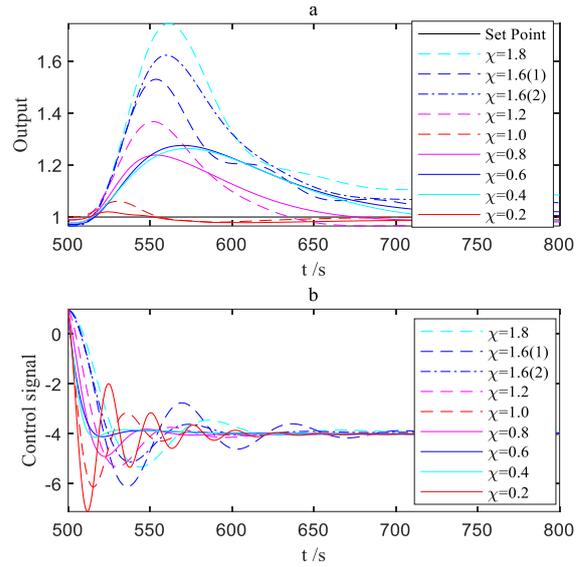

Fig. 11. The local enlarged drawing of the disturbance rejection in Fig. 9. (a: the output response, b: the control signal)

## 6 CONCLUSIONS

To handle the control difficulties caused by high-order dynamics, a control structure based on fractional order [PI] controller and fractional order Smith-like predictor for a class of high order systems in the type of $K/(Ts+1)^n$ is proposed in this paper. The analysis of the tracking and disturbance rejection is illustrated based on the terminal value theorem and shows that the proposed control structure can ensure that the closed-loop system converges to the set point without static error and the closed-loop system recovers to its original state when the input disturbance occurs. Then, the simulations about the influence on the control performance and control signal with different $\chi$ are discussed and the results show that the control performance can be improved and the energy of the control signal can be reduced simultaneously when the order $\chi$ of the proposed structure is chosen no more than one. This can verify that the fractional order Smith-like predictor with has an advantage over that of the integral order Smith-like predictor with $\chi=1$. What is more, how to select an appropriate $\chi$ will be the next work.